\documentclass[3p,times,procedia]{elsarticle}
\usepackage{nupha_ecrc}
\usepackage{hyperref}

\volume{00}

\firstpage{1}

\journalname{Nuclear Physics A}

\runauth{}

\jid{nupha}

\jnltitlelogo{Nuclear Physics A}

\usepackage{amssymb}
\usepackage{amsmath}

\usepackage[figuresright]{rotating}

\usepackage{bm}
\def\x{{\bm x}}
\def\p{{\bm p}}
\usepackage{tikz}
\tikzstyle{myrectangle}=[rectangle,rounded corners, fill=white, 
draw=uibred,align=center,line width=1pt]
\tikzstyle{mycircle}=[fill=white, fill opacity = 0.8, draw, line width=1pt]
\tikzstyle{myfigure}=[anchor=south west,fill=white]
\definecolor{uibred}{RGB}{167, 38, 47}

\def\ubr#1#2{\underbrace{#1}_{\text{#2}}}

\def\Fig#1{Fig.~\ref{#1}}
\begin{document}

\begin{frontmatter}

\dochead{XXVIth International Conference on Ultrarelativistic Nucleus-Nucleus 
Collisions\\ (Quark Matter 2017)}

\title{Initial conditions for hydrodynamics from
kinetic theory equilibration}

\author[c,d]{Aleksi Kurkela}

\address[c]{
Theoretical Physics Department, CERN, Geneva, Switzerland
}
\address[d]{Faculty of Science and Technology, University of Stavanger, 4036 Stavanger, Norway}

\author[a,b]{Aleksas Mazeliauskas}
\address[a]{Department of Physics and Astronomy, Stony Brook University, Stony 
Brook, New York 11794, USA}
\address[b]{Institut f\"{u}r Theoretische Physik, Universit\"{a}t Heidelberg, 
69120 Heidelberg, Germany}
\ead{a.mazeliauskas@thphys.uni-heidelberg.de}
\author[a]{Jean-Fran\c cois Paquet}
\author[f]{S\"oren~Schlichting}
\address[f]{Department of Physics, University of Washington, Seattle, WA 98195-1560, USA}
\author[a]{Derek Teaney}

\begin{abstract}
We use effective kinetic theory to study the pre-equilibrium dynamics in heavy-ion 
collisions. We describe the evolution of linearized energy perturbations on top of out\nobreakdash-of\nobreakdash-equilibrium 
background to the 
energy-momentum tensor at a time when hydrodynamics becomes applicable. We 
apply this description to IP-Glasma initial conditions and find an overall smooth 
transition to hydrodynamics. In a phenomenologically favorable range of 
$\eta/s$ values, early time dynamics can be accurately described in terms of a few functions of a scaled time variable $\tau T/(\eta/s)$. Our framework can be readily applied to other initial state models to provide the pre-equilibrium dynamics of the energy momentum tensor.
\end{abstract}


\end{frontmatter}

\section{Introduction}
\label{}

The expansion of Quark Gluon Plasma fireball in heavy ion collisions is 
successfully described by relativistic hydrodynamics with small shear viscosity 
over entropy ratio 
$\eta/s$~\cite{Heinz:2013th,Luzum:2013yya,Teaney:2009qa}. %
However, the early time equilibration and isotropization necessary for this
hydrodynamic description is outside the scope of hydrodynamics and 
initial conditions at hydrodynamic initialization time 
$\tau_\text{init}\sim 1\,\text{fm}$ have to be supplied by other 
models. A desirable pre-equilibrium description would 
naturally and 
smoothly transition to hydrodynamics and, therefore, eliminate the dependence 
on the initialization time 
$\tau_\text{init}$~\cite{vanderSchee:2013pia,Romatschke:2015gxa,Kurkela:2016vts}.

In the limit of weak coupling at high collision energies, the 
early time dynamics can be described by a combination of Color-Glass Condensate (CGC) 
saturation framework~\cite{Iancu:2002xk,Iancu:2003xm,Gelis:2010nm} and 
effective kinetic theory~\cite{Baier:2000sb,Arnold:2002zm}.  Classical 
lattice simulations showed that the so called ``bottom-up" is the preferred 
thermalization scenario in the weak coupling 
limit~\cite{Berges:2013fga,Berges:2013eia}, and kinetic theory realization of 
uniform background evolution at 
moderate values of the 
coupling constant (which determines the effective $\eta/s$) 
reaches hydrodynamic behavior in a
phenomenologically reasonable time~\cite{Kurkela:2015qoa}. In 
this work we 
describe a practical implementation of kinetic theory pre-equilibration stage 
for the transverse energy and momentum 
perturbations~\cite{Keegan:2016cpi,Kurkela:2017a}.

\section{Kinetic theory response}
\begin{figure}
\centering
\begin{tikzpicture}[>=stealth]
\node[anchor=south west] at (0,0) (image) 
{\includegraphics[width=0.35\linewidth]{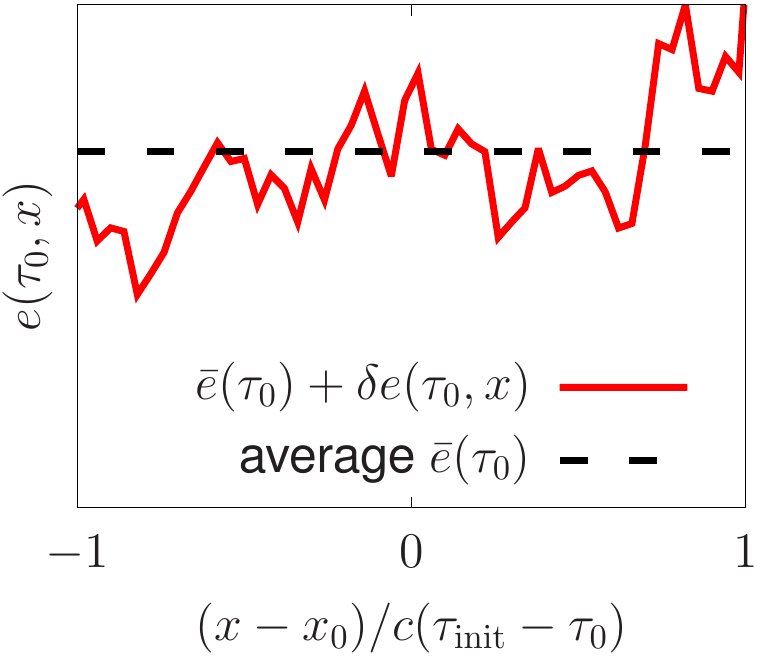}};
\node[anchor=west,xshift=0.5cm] (image2) at (image.east) 
{\includegraphics[width=0.45\linewidth]{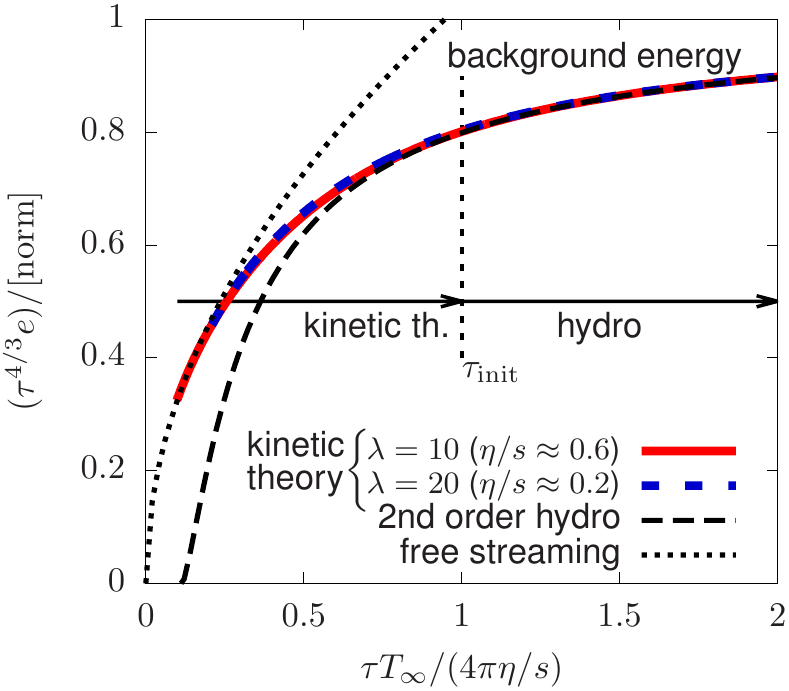}};
\node at (0,0) {a)};
\node at (6.5,0) {b)};
\end{tikzpicture}
\caption{\label{fig2}(a) Energy density decomposition to the average 
and perturbations within a causal 
circle 
$|\x-\x_0|<c(\tau_\text{init}-\tau_0)$, which  determines the system response 
at 
$(\tau_\text{init},\x_0)$. 
(b) Kinetic theory 
equilibration 
of boost invariant background 
energy density in scaled time units $\tau T_\infty/(4\pi\eta/s)$, where 
$T_\infty(\tau)\equiv 1/\tau^{1/3} \lim_{\tau\rightarrow \infty} 
(\tau^{1/3}T(\tau))$. At early times the expansion resembles free streaming, 
but at late times the evolution agrees with the hydrodynamic gradient expansion 
given in 
Eq. \eqref{hydro}}
\end{figure}

One of the key features of equilibration is the memory loss about the 
initial state. The late time hydrodynamic evolution of heavy ion collisions is 
given in terms of
the conserved charges (energy and transverse momentum), therefore we use the 
energy momentum tensor components 
$T^{\mu\nu}$, 
i.e.\ the first moments of the distribution function, to 
characterize the out-of-equilibrium evolution of  particle distribution 
function $f(\p,\x,\tau)$ in the effective kinetic theory. 
Causality restricts the system response to its neighborhood, and the time when 
hydrodynamic description becomes applicable $\tau_\text{init}$ is 
typically short compared 
to the size of heavy ion collision area. Therefore we use linearized 
perturbations of the distribution function to calculate the  out-of-equilibrium 
energy momentum 
tensor perturbations 
$\delta 
T^{\rho\sigma}(\x, \tau)$.
Then the total energy momentum tensor at $\tau_\text{init}$  can be written as 
a sum of background $T^{\mu\nu}_\text{avg}(\x, 
\tau_\text{init})$ and the convolution of kinetic theory response function 
$G_{\rho\sigma}^{\mu\nu}(\x-\x',\tau_0,\tau_\text{init})$ to initial state 
perturbations $\delta 
T^{\rho\sigma}(\x', \tau_0)$ \footnote{ In Eq.~\eqref{one} the 
indexes of 
$\delta T^{\rho\sigma}$ 
refer to the conserved components of energy momentum tensor, i.e. $\delta 
e=\delta T^{00}$ 
and 
$g^i=\delta T^{0i}$.}
\begin{equation}
T^{\mu\nu}(\x , \tau_\text{init}) =T^{\mu\nu}_\text{avg}(\x, \tau_\text{init}) 
+ \int_{|\x-\x'|<|\tau_\text{init}-\tau_0|}\!\!\!\! d^2\x' 
G_{\rho\sigma}^{\mu\nu}(\x-\x',\tau_0,\tau_\text{init})\,\delta 
T^{\rho\sigma}(\x', \tau_0).\label{one}
\end{equation}
Here the background $T^{\mu\nu}_\text{avg}(\x, \tau_\text{init})$ is taken to 
be uniform and boost invariant within the causal circle (see \Fig{one}(a)). The 
nonlinear 
background equilibration is described by kinetic theory map  $\mathcal{F}$
\begin{equation}
T^{\mu\nu}_\text{avg}(\x, \tau_\text{init}) 
=\mathcal{F}\left[T^{\rho\sigma}_\text{avg}(\x, 
\tau_0),\tau_0, \tau_\text{ini}\right]\label{two},
\end{equation}
which is computed by direct simulation. Based on a suitable form of 
perturbations of the quasi-particle distribution function 
$f(\x,\p,\tau)$, the coordinate Green 
functions
$G_{\rho\sigma}^{\mu\nu}(\x-\x',\tau_0,\tau_\text{init})$ are
calculated using the linearized evolution of the Boltzmann equation with 
gluonic elastic scatterings and inelastic particle number changing 
processes~\cite{Keegan:2016cpi,Kurkela:2017a}.

The expected late time behavior of the boost invariant energy density is given 
by a 
hydrodynamic
gradient expansion
\begin{equation}
{\text{Hydro 
prediction:}\quad\tau^{4/3}\bar{e}}={(\tau^{4/3}\bar{e})_\infty}\bigg(\ubr{\vphantom{C_2\bigg(\frac{\eta/s}{\tau
 T}\bigg)^2}1}{ideal}-\ubr{\vphantom{C_2\bigg(\frac{\eta/s}{\tau 
 T}\bigg)^2}\frac{8}{3}\frac{\eta/s}{\tau
T}}{viscous}+\ubr{C_2\bigg(\frac{\eta/s}{\tau T}\bigg)^2}{2nd order 
hydro}+\ldots\bigg)\label{hydro}
\end{equation}
where $C_2$ is a constant depending on the second order transport coefficients. 
In 
\Fig{one}(b) we compare the background energy evolution in kinetic theory to a 
second order hydrodynamic asymptotics. 
Empirically we find that for the relevant range of $\eta/s$ values, 
equilibration is universal if plotted in units of kinetic theory relaxation 
time $\tau_R = {(\eta/s)}/{T}$.
Similarly to the background, kinetic theory response functions at the same 
$\tau/\tau_R$ look 
the same if 
plotted as a function of radial distance $r/\tau$ in the causal circle (not 
shown).

\section{Results and Conclusions}

\begin{figure}
\centering
\begin{tikzpicture}
\node[anchor=south west] (image) at (0,0) {
\includegraphics[width=0.45\linewidth]{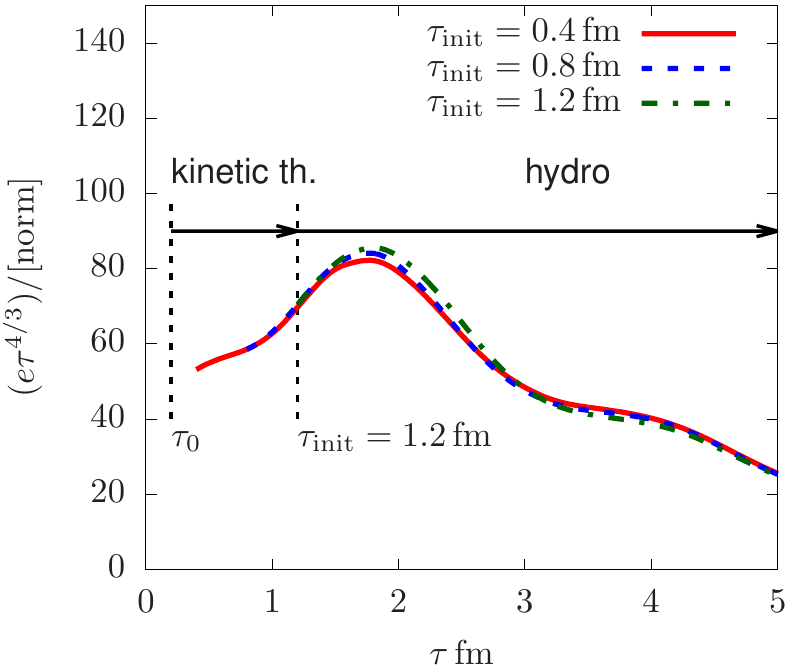}\quad\quad
\includegraphics[width=0.45\linewidth]{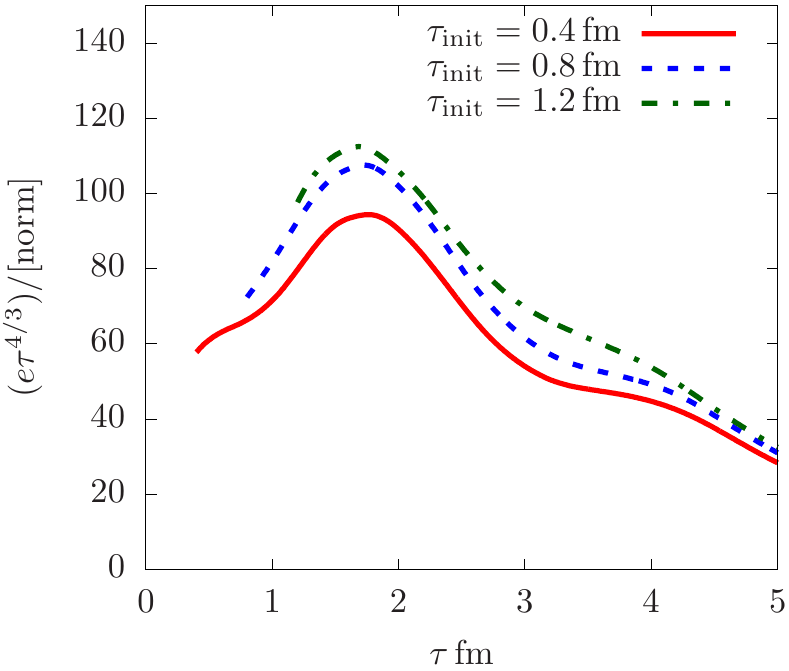}};
\begin{scope}[x={(image.south east)},y={(image.north west)}]
\node[myrectangle] at (0.25,0.25) {kinetic theory};
\node[myrectangle] at (0.75,0.25) {free streaming};
\end{scope}
\node at (0.5,0.5) {a)};
\node at (7.8,0.5) {b)};
\end{tikzpicture}
\caption{\label{fig3}
Energy density evolution at the center 
of a fireball after matching to hydrodynamics $\tau_\text{init}=0.4,0.8,\text{ 
and }1.2\,\text{fm}$. The pre-equilibrium evolution 
$\tau_0<\tau<\tau_\text{init}$ done in
(a) kinetic theory  (b) free streaming.}
\end{figure}
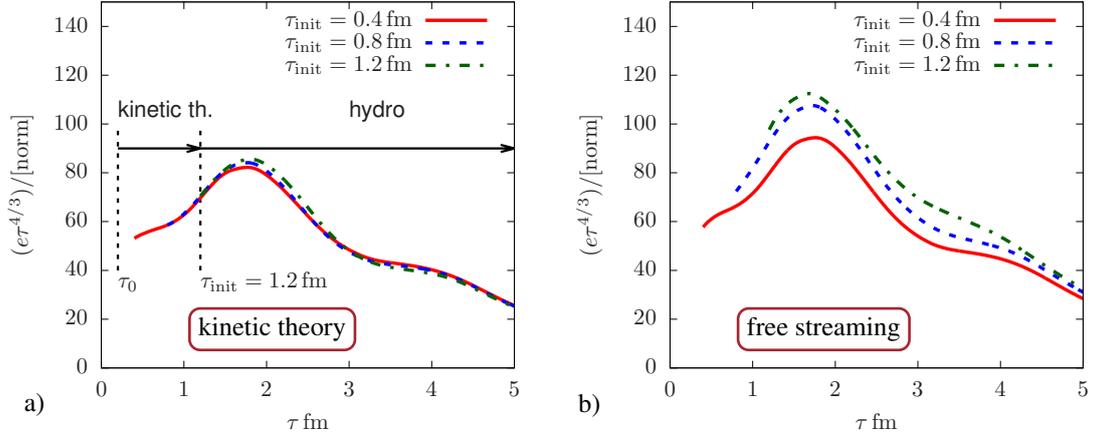

\begin{figure}
\centering
\includegraphics[width=\linewidth]{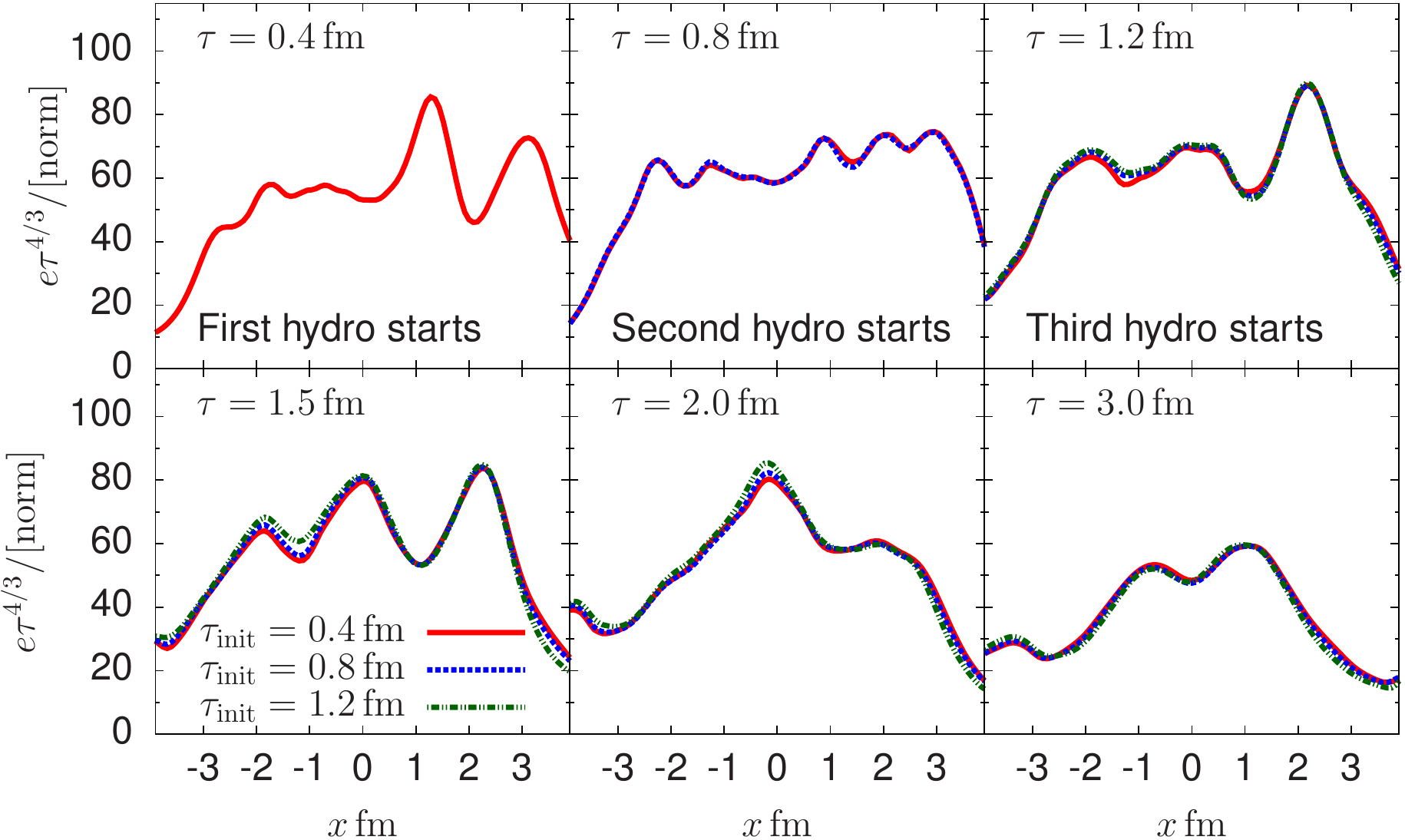}
\caption{\label{fig4} Transverse slice of energy density after matching to 
hydrodynamics $\tau_\text{init}=0.4,0.8,\text{ 
and }1.2\,\text{fm}$ with  the kinetic theory pre-equilibrium 
evolution.}
\end{figure}

We apply kinetic theory pre-equilibrium evolution to a realistic energy density 
profile taken from IP\nobreakdash-Glasma initial state model at 
$\tau_0=0.2\,\text{fm}$~\cite{Schenke:2012wb,Schenke:2012fw}. 
The energy perturbations and the background are then propagated to hydrodynamic
initialization time $\tau_\text{init}$, which is varied.  In \Fig{fig3} we 
show the energy density evolution at the center 
of the fireball after matching to hydrodynamics at 
$\tau_\text{init}=0.4,0.8,\text{ and }1.2\,\text{fm}$. We observe that 
the kinetic theory pre-equilibrium smoothly matches 
hydrodynamic evolution and the 
subsequent evolution is largely independent of the switching time $\tau_\text{init}$. 
In contrast,  the free streaming evolution does not match hydrodynamics and the 
late time behavior is sensitive to hydrodynamic initialization time.
In \Fig{fig4} we show the transverse slice of energy density of the same event 
at 
different times $\tau$. We see that the transverse profile agrees well between 
all three hydrodynamic initializations. During the pre-equilibrium evolution 
the initial energy perturbations also induces a transverse flow, which is an 
important input in hydrodynamic initialization and can be captured by kinetic 
theory 
evolution.

Based on a microscopic description of bottom-up thermalization, we demonstrated 
the feasibility of event-by-event simulations of the pre-equilibrium dynamics 
for realistic initial conditions. Naturally our description can be smoothly 
matched to the subsequent hydrodynamic evolution, thus eliminating the 
dependence on hydrodynamic initialization time.

\subsection*{Acknowledgments}

The authors would like to thank Liam Keegan and Bj\"orn 
Schenke for their contributions at 
the beginning of this work.
Results in this paper were obtained using the high-performance computing system 
at the Institute for Advanced Computational Science at Stony Brook University. 
This work was supported in part by the U.S. Department of Energy, Office of 
Science, Office of Nuclear Physics  under Award Numbers 
DE\nobreakdash-FG02\nobreakdash-88ER40388 (A.M., J.-F.P., D.T.) and 
DE-FG02-97ER41014 (S.S.). A.M. would like to thank the organizers of QM 
2017 for the student support.

\bibliographystyle{elsarticle-num}
\bibliography{master.bib}

\end{document}